# Graphical Abstract

**Impact of Spinning Droplets onto Superhydrophobic Surfaces: Asymmetric Tumbling Rapid Rebound**

Jinyang Wang, Feifei Jia, Xiaoyun Peng, Peng Zhang, Kai Sun, and Tianyou Wang

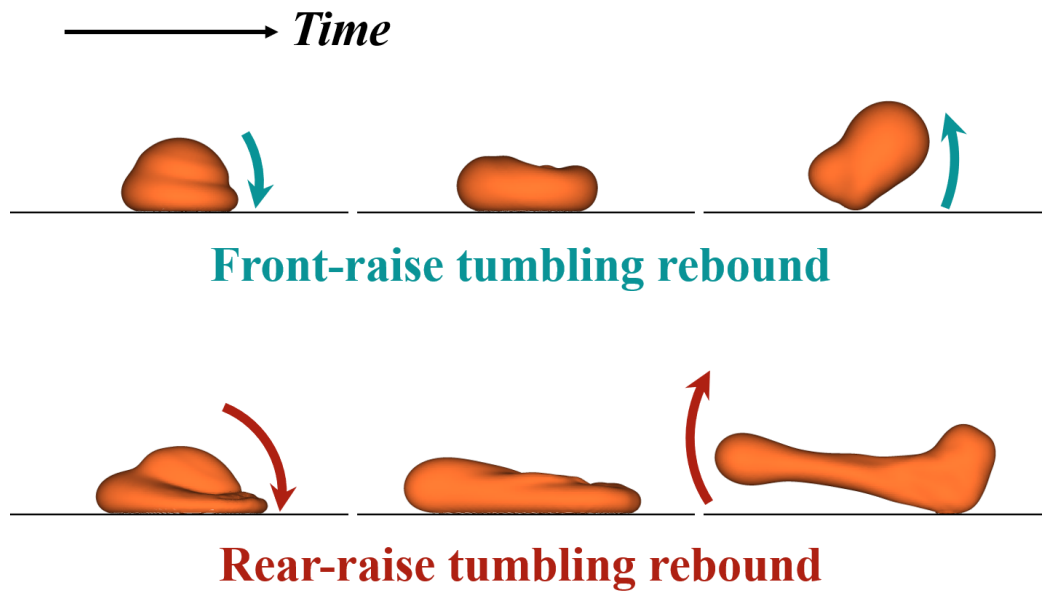

# Highlights

**Impact of Spinning Droplets onto Superhydrophobic Surfaces: Asymmetric Tumbling Rapid Rebound**

Jinyang Wang, Feifei Jia, Xiaoyun Peng, Peng Zhang, Kai Sun, and Tianyou Wang

- The spinning motion of droplets leads to two novel rebound scenarios: front-raise tumbling rebound and rear-raise tumbling rebound.

- The reversed torque from inertial impact may cause the droplet's detachment angular momentum to oppose the observed tumbling direction.

- A unified theoretical model was established to explain how $We$ and $\Omega$ enhance asymmetric deformation and rapid rebound of spinning droplets.

# Impact of Spinning Droplets onto Superhydrophobic Surfaces: Asymmetric Tumbling Rapid Rebound

Jinyang Wang[a,b], Feifei Jia[a], Xiaoyun Peng[a], Peng Zhang[b,*], Kai Sun[a,*], and Tianyou Wang[a]

[a]*State Key Laboratory of Engines, Tianjin University, Tianjin, 300350, China*
[b]*Department of Mechanical Engineering, City University of Hong Kong, Hong Kong, 999077, China*

**Abstract**

The impact dynamics of spinning droplets onto superhydrophobic surfaces was studied by using Volume-of-Fluid simulations, covering broad ranges of *Weber* number ($We$) and dimensionless angular velocity ($\Omega$). The computational results were validated by high-speed imaging experiments, with particular focus on the types of rebound, asymmetric deformation, and droplet-wall contact time. Results show that, the spinning motion of droplets leads to two novel rebound scenarios. Specifically, the front-raise tumbling rebound occurs at a lower $\Omega$ and is caused by the unsymmetrical Laplace pressure, while the rear-raise tumbling rebound emerges at a higher $\Omega$ and is attributed to the rotational inertia. The angular momentum of the spinning droplet is dissipated or even reversed, while its direction upon detachment is inconsistent with the visually observed spinning motion. With the increase of the angular velocity, the droplet-wall contact time is largely reduced, which is attributed to the asymmetric spreading by the spinning motion rather than the increased kinetic energy. A theoretical model was also established to predict asymmetric spreading and the contact time and validated against numerical results in wide ranges of $We$ and $\Omega$.

*Corresponding author
  *Email addresses:* `penzhang@cityu.edu.hk` (Peng Zhang), `sunkai@tju.edu.cn` (Kai Sun)

1. Introduction

Droplets impact on solid surfaces is ubiquitous in nature and holds practical significance for many industrial processes such as spray cooling [1], engine combustion [2] inkjet printing [3], *etc.*. Various outcomes of droplet impact have been identified, including deposition, prompt splashing, corona splashing, receding breakup, rebound, and partial rebound [4, 5, 6, 7, 8]. Recently, increasing attention has been paid to superhydrophobic surfaces, which exhibit excellent non-wetting properties and thus hold great potential in self-cleaning [9], anti-icing/frosting [10, 11], anti-corrosion [12], and phase-change heat transfer [13].

Most previous studies have focused on the simplified problem of droplets impacting normally onto superhydrophobic surfaces. For the conventional type of droplet rebound, the droplet (with an initial diameter $D_0$) firstly spreads to the maximum diameter $D_m$ with a minimum thickness $h$, then retracts and ultimately detaches from the surface. The droplet-wall contact time, from initial contact to final detachment, was commonly used to evaluate the speed of droplet rebound. Within the range of $We$ numbers in the inertial regime, the maximum spreading ratio of the droplet follows $D^* = D_m/D_0 \sim We^{1/4}$ due to the balance between capillary force and inertia force [14]. The contact time was generally found to scale as the inertial-capillary time $t_\sigma \sim \sqrt{\rho_1 R_0^3/\sigma}$ that being independent of the impact velocity [15] Such a theoretical limit could be broken through by manipulating the droplet impact process. Specifically, Liu *et al.* [16] patterned the superhydrophobic surface with submillimetre-posts and nanotextures and discovered a new type of rebound termed "pancake bouncing", in which the droplet quickly rebounds in a pancake-like shape without retraction. Bird *et al.* [17] introduced a macroscopic ridge structure on the surface, showing obviously shortened non-axisymmetric recoil of the spread film. Liu *et al.* [18] studied drop impact dynamics on a concave surface, and the asymmetric bouncing leads to 40% reduction in the contact time. Liu *et al.* [19] designed a superhydrophobic surface with stepped structures and simultaneously achieved directional bouncing and reduction of the contact time. Regarding the droplet properties, the influence of droplet shape [20, 21], composition [22, 23, 24], viscosity [25, 26], and non-Newtonian effects [27] were also discussed.

In real-world scenarios, the relative motion between the droplet and the surface is more complicated, and normal impact does not necessarily occur. If the surface is inclined [28, 29] or moving [30], the droplet would impact the surface obliquely [31]. The impact angle was found to significantly affect the impact dynamics due to symmetry breaking. Šikalo et al. [32] experimentally found that droplets tend to deposit on the surface at high impact angles, whereas low impact angles are more prone to rebound. Experiments of Antonini et al. [33] demonstrated that surface inclination enhances droplet rebound, reducing the contact time by up to 40%. Studies also found a variety of new rebound types with reduced contact time [29, 34]. Yin et al. [35] found that oblique impact tends to facilitate the occurrence of incomplete-retracting bouncing and tumbling bouncing, thereby significantly reducing the contact time. Zhan et al. [30] studied the impact of droplets onto superhydrophobic surfaces with horizontal motion. The relative motion generated additional viscous forces in the air film, leading to asymmetric droplet deformation and a reduction in contact time. Tao et al. [36] observed that droplets impacting the center of rotating superhydrophobic surfaces could detach rapidly in a "doughnut" shape without retraction, reducing the contact time by about 40% compared to stationary surfaces.

In addition, vibration as another form of relative motion, was found to induce various oscillation modes of different characteristic frequencies [37], thereby serving as an effective method for removing droplets from superhydrophobic surfaces [38]. Droplets deposited on an inclined surface may either climb or slide due to vertical vibrations of the substrate [26, 39]. Directional movement of droplets can also be achieved through the superposition of horizontal and vertical vibrations of the substrate [40]. Wang et al. [41] found that the threshold for droplet detachment from textured vibrating surfaces depends on the droplet size and the maximum apparent droplet-wall contact area. Sun et al. [38] developed a theoretical model to determine the most efficient conditions for droplet removal from vibrating superhydrophobic surfaces. Their results show that, the droplet energy must exceed the surface adhesion energy to detach the droplets from the surface, which can be achieved efficiently by vibrating the surface at its resonant frequency. Böhringer et al. [42] introduced textured ratchets to induce asymmetric pinning forces along the three-phase contact line to control droplet movement under vertical vibration.

Inspired by the three types of molecular motion [43], namely translation, vibration, and rotation, we note that droplets may also possess rotational motion, which are extensively observed in various scenarios such as coalesced droplets after off-center collision [44, 45, 46, 47], rebound droplets upon obliquely impacting on surfaces [48, 49], and droplets in complex flow fields [50, 51] or under external forces [52, 53]. Among the very limited research, Hill *et al.* [52] studied the non-axisymmetric shapes of a magnetically levitated and spinning water droplet, and they identified the equatorial traveling waves to give the droplet threefold, fourfold, and fivefold symmetry. Melo *et al.* [54] studied the fingering instability during the spreading of a spinning drop. He *et al.* [55, 56] studied the dynamic of spinning droplet collisions, revealing that the non-axisymmetric flow enhances mass mixing and yields varied collision outcomes.

Albeit of these worthy advances, however, the impact of spinning droplets on surfaces has been rarely studied and remains inadequately understood. In this work, we aim to investigate the effect of rotation on the dynamics of droplet impact on superhydrophobic surfaces. Both experiments and simulations are performed, with particular attention paid to the droplet spreading and rebound process, as well as their effects on the contact time. In the following, experimental and computational methods, results and discussion, and concluding remarks will be presented from Section 2 to Section 4.

## 2. Experimental and Computational methods

### 2.1. Experimental method

The schematic diagram of the experimental setup is shown in Fig. 1. Deionized water droplets with tracer particles were generated by a syringe on an inclined, V-shaped slide track. Under the influence of gravity, the droplets acquired spinning motion along the superhydrophobic slide track and eventually impacted on the superhydrophobic surface vertically. Such methods for generating spinning droplets are effective, as previous studies [57, 58] have reported significant rolling of droplets on superhydrophobic surfaces. The V-shaped design ensured that the droplet rotation remained in-plane, which was further confirmed by the experimental observation that the droplet and its motion consistently stayed within the focal plane. The normal impact velocity was characterized by monitoring droplet's center of mass. The angular velocity was determined by the Particle Tracking Velocimetry (PTV) experiments, which track the motion of tracer particles dispersed in the droplets [57, 59, 60, 61]. The difference between the velocities of droplet center and particle was used to calculate the rotational velocity of droplet. In this study, the tracer particles were Polyamide (($C_6H_nNO)_n$) with mean diameters ranging from 37 to 75 $\mu$m, which was sufficiently small to passively follow the flow pattern yet large enough to avoid the Brownian motion. According to the Stokes' drag law, the response time of the particles can be expressed as $\tau_\mathrm{p} = d_\mathbf{p}^2 \rho_\mathrm{p}/(18\mu_1)$, where $d_\mathbf{p}$ is the diameter of the particle, $\rho_\mathrm{P}$ is the particle density (about 1 g/cm$^3$), $\mu_1$ is the dynamic viscosity of water. The particle *Stokes* number is: $St = \tau_\mathrm{p}/\tau_\mathrm{F}$, where $\tau_\mathrm{p}$ denotes the characteristic time in the flow. For spinning droplets impacting on superhydrophobic surfaces in this study, $\tau_\mathrm{F}$ can be regarded as the inertial-capillary time $t_\sigma$. Therefore, $St$ is on the order of $O(10^{-11})$, indicating that the tracer particles were able to follow the flow with sufficient fidelity [62]. Experimental observations further confirmed that the particles had no appreciable influence on the impact dynamics. The superhydrophobic slide track and superhydrophobic surface were fabricated by spraying and drying the superhydrophobic coating, which was a suspension of adhesive/fluorinated silica core/shell microspheres. The superhydrophobic coating has been proven to have excellent impalement resistance, mechanical robustness, and weather resistance simultaneously [63]. The scanning electron microscopy (SEM) image in Fig. 1(a) shows that the nanoparticles formed a loosely porous layer, which imparted hydrophobicity to the surface and yielded an equilibrium contact angle of 160°, as illustrated in Fig. 1(b).

The experiments were conducted at room temperature (22 ±2°C) and atmospheric pressure, with the droplet impact process being varied by adjusting the slide track angle $\alpha$, the initial position, and the size of the droplet to change both the angular velocity and the impact $We$ number. To ensure an ideal normal impact of the spinning droplet, the V-shaped slide track was designed to be axisymmetric and kept perpendicular to the surface, with a short distance between the track outlet and the surface to neglect the influence of gravity on the impact angle. The droplet motion in the $x$-$z$ plane was captured in detail by using a high-speed camera (Photron FASTCAM SA1.1) and a macro lens (Tokina 100 mm f/2.8D) positioned along the y-axis. The camera operated at 5400 frames per second (fps) with a resolution of 1024 × 1024 pixels. A LED lamp (Hecho S5000, 60-w) was used in combination with a diffuser and reflector to provide front-side illumination for high-speed imaging. The angular velocities of the droplets were calculated based on the positional changes of tracer particles through image processing, while the translational velocities of the droplets were determined by the movement of their centroids.

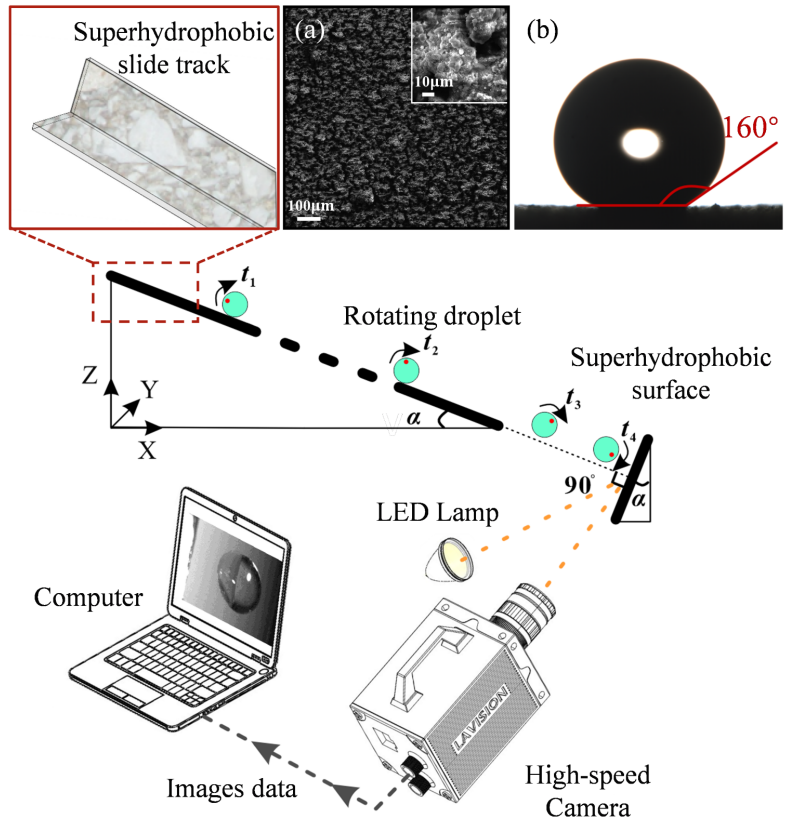

Figure 1: Schematic diagram of the experimental setup and the superhydrophobic surface: (a) Scanning electron microscopy image of the superhydrophobic surface. (b) Measured contact angle of the superhydrophobic surface.

## 2.2. Computational method

The finite-volume-method-based partial differential equation solver, Basilisk C [64], was employed as the numerical platform. The octree space discretization was used for adaptive-mesh-refinement (AMR) [65, 66, 67], which dynamically changes the grid structure based on the local volume fraction and the velocity field. This approach is able to significantly reduce the computational effort while maintaining the accuracy [68, 69, 70].

### 2.2.1. Volume of Fluid Method

The present simulation was conducted with the volume of fluid (VOF) method. The governing equations include the continuity equation, the momentum equation, and the advection equation:

$$\nabla \cdot \mathbf{u} = 0, \tag{1}$$

$$\rho(\frac{\partial \mathbf{u}}{\partial t} + \mathbf{u} \cdot \nabla \mathbf{u}) = -\nabla p + \nabla \cdot (2\mu \boldsymbol{D}) + \mathbf{T}_\sigma, \tag{2}$$

$$\frac{\partial f(\mathbf{x})}{\partial t} + \nabla \cdot (\mathbf{u} f(\mathbf{x})) = 0, \tag{3}$$

where $\mathbf{u}$ represents the velocity vector, $\rho$ represents the density, $p$ represents the pressure, $\mu$ represents the dynamic viscosity, and $\boldsymbol{D}$ represents the deformation tensor defined as $(\nabla \mathbf{u} + \nabla \mathbf{u}^{\mathrm{T}})/2$. In Eq. (2), $\mathbf{T}_\sigma$ is the surface tension term modeled as $\sigma \kappa \mathbf{n} \delta_{\mathrm{s}}$ [71], where $\sigma$ is the liquid-gas surface tension coefficient, $\kappa$ is the local curvature, $\mathbf{n}$ denotes the normal vector of the interface, and $\delta_s$ is the Dirac delta function that is non-zero exclusively at the liquid-gas interface. In Eq. (3), $f(\mathbf{x})$ represents the volume fraction of the liquid phase in the control volume, $\mathbf{x} = (x, y, z)$ represents the space coordinate. Thus, $f(\mathbf{x}) = 1$ represents pure liquid, $f(\mathbf{x}) = 0$ represents pure gas, and $0 < f(\mathbf{x}) < 1$ represents the interface region. The density and viscosity of the fluid were determined by:

$$\rho = \rho_{\mathrm{l}} + \rho_{\mathrm{g}}(1 - f(\mathbf{x})), \tag{4}$$

$$\mu = \mu_{\mathrm{l}} + \mu_{\mathrm{g}}(1 - f(\mathbf{x})), \tag{5}$$

where the subscripts l and g denote the liquid phase and the gas phase, respectively.

*2.2.2. Computational setup*

The schematic diagram of the three-dimensional (3-D) computational domain is shown in Fig. 2. The length of the cubic domain $L$ was substantially larger than the droplet diameter ($L=15R_0$) to eliminate possible boundary effects. Initially, the droplet of radius $R_0$ was placed above the superhydrophobic surface with a normal velocity of $V_0$ and a spinning angular velocity $\omega_0$ . The ambient environment was quiescent air. The non-slip boundary condition was applied to the bottom solid wall, while the other side/top walls were specified as open boundaries. To simplify the simulations, the wall was assumed to be fully superhydrophobic, namely a thin air layer always exists between the droplet and the surface, which has been adopted as an efficient approach in similar problems by Loshe and his collaborators [72, 73]. Such an assumption can be supported by a few experiments [74, 72, 75], which found that air layers were commonly present during the impact of droplets on both wet and non-wet surfaces, except for those destabilizing and collapsing during high-speed splashing cases. This is primarily due to the relatively low adhesion work between the droplet and the superhydrophobic surface wall. According to the classical Young-Dupré equation [76], the adhesion work per unit area is expressed by $\Delta W_{\rm sl} = \sigma(1 + \cos\theta)$, where $\theta$ is the contact angle. For droplets with a millimeter-scale radius on superhydrophobic surfaces ($\theta \geq 150°$), the adhesive force is generally on the order of $F_{\rm ad} \sim O(10^{-6})$ N, which is negligible compared to the other dominant forces such as surface tension, inertia, and viscous force. To refine the grid resolution, the droplet interface and surrounding regions were refined by using a wavelet-based error method [77], which is already implemented in Basilisk.

According to the scaling analysis, five dimensionless parameters were selected to describe the flow similarity of the droplet impact process, including the normal *Weber* $We = \rho_1 R_0 V_0^2/\sigma$, the dimensionless angular velocity $\Omega = \omega_0 t_\sigma = \omega_0 \sqrt{\rho_1 R_0^3/\sigma}$, the *Ohnesorge* number $Oh = \mu_1/\sqrt{\rho_1 R_0 \sigma}$, the liquid-gas density ratio $\rho_1/\rho_{\rm g}$, and the liquid-gas viscosity ratio $\mu_1/\mu_{\rm g}$. In the present water-air system, $\rho_1/\rho_{\rm g}$ and $\mu_1/\mu_{\rm g}$ were fixed at 813 and 55, respectively. Throughout the experiments, the droplet radius $R_0$ varied within a narrow range of 1.3 ±0.2 mm, and thus $Oh$ was approximately fixed at 0.003. Therefore, among the five dimensionless parameters, $We$ and $\Omega$ were chosen as the controlling variables. These dimensionless parameters are summarized in Table 1. In the following, time is normalized by the inertial-capillary time scale as $T = t/t_\sigma$, with $T=0$ denoting the first instant of droplet-wall contact.

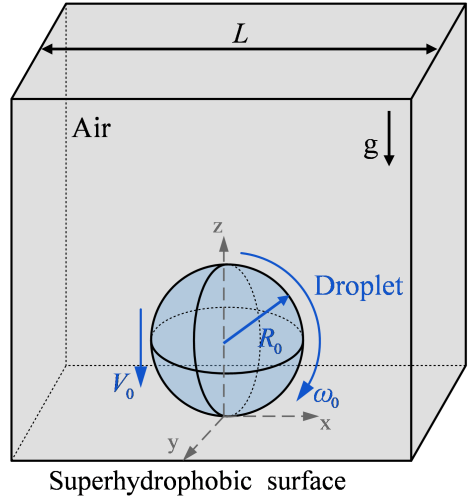

Figure 2: Schematic of the computational domain and the initial condition.

| Dimensionless parameters | Value/Range |
|---|---|
| *Weber* number, $We = \rho_l R_0 V_0^2/\sigma$ | $1 \sim 10$ |
| Dimensionless angular velocity, $\Omega = \omega_0 \sqrt{\rho_l R_0^3/\sigma}$ | $0 \sim 5$ |
| *Ohnesorge* number, $Oh = \mu_l/\sqrt{\rho_l R_0 \sigma}$ | 0.003 |
| Liquid-gas density ratio, $\rho_l/\rho_g$ | 813 |
| Liquid-gas viscosity ratio, $\mu_l/\mu_g$ | 55 |

Table 1: Dimensionless parameters in the present study.

## 3. Results and discussion

*3.1. Experimental validation and grid independence analysis*

Fig. 3 shows snapshots of typical normal and spinning droplet impact processes. Both experimental results (top row) and numerical results (bottom row) are presented. The experimental snapshots of normal impact in Fig. 3(a) were captured by using a backlighting setup, ensuring precise and clear visualization of the droplet interface evolution. As shown in Fig. 3(a), during the early stage of normal impact with $We = 1.1$ ($T = 0 \sim 0.74$), the droplet undergoes symmetric deformation and spreading. Driven by the inertia, the droplet's kinetic energy transitions into surface energy, reaching its maximum spreading diameter at $T = 0.99$. Then, the large capillary force along the droplet edge induces retraction motion ($T = 1.36 \sim 1.73$), during which the surface energy converts back to the kinetic energy, ultimately leading to complete detachment from the surface at $T = 2.35$. Fig. 3(b), (c), and (d) illustrate the spinning impact processes with increasing $We$ and $\Omega$. The experimental results (top row) in these cases employed front lighting to capture the motion of tracer particles within droplets. Compared to normal impact, the addition of angular velocity breaks the symmetry of droplet dynamics, such as shifting the directions of spreading and retraction, and rebounding obliquely from the superhydrophobic surface. Such asymmetry intensifies with increasing $We$ (from 1.2 to 2.7) and $\Omega$ (from 0.4 to 0.65), as a tilted tip can be observed shortly after impact (see $T = 0.67$ in Fig. 3(c) and $T = 0.70$ in Fig. 3(d)). Overall, the experimental and numerical results show good agreement. The further quantitative validation and the grid independence test of above four cases are presented in Fig.4 by comparing the trajectories of the droplet apex. The height of the droplet $h$ was normalized by $h^* = h/2R_0$ The experimental error is estimated to be $\pm 5\%$ of the measured values. Grid independence test was performed by comparing the simulation results with different grid resolutions of $N^* = R_0/\Delta = 9, 17, 34$, and 68, where $\Delta$ represents the minimum grid size in the computational domain. Results show that, when $N^*$ exceeds 34, the simulated $h^*$ aligns well with the experimental results. Therefore, in view of the balance between the computational accuracy and efficiency, a grid resolution of $N^* = 34$ was utilized in the present study.

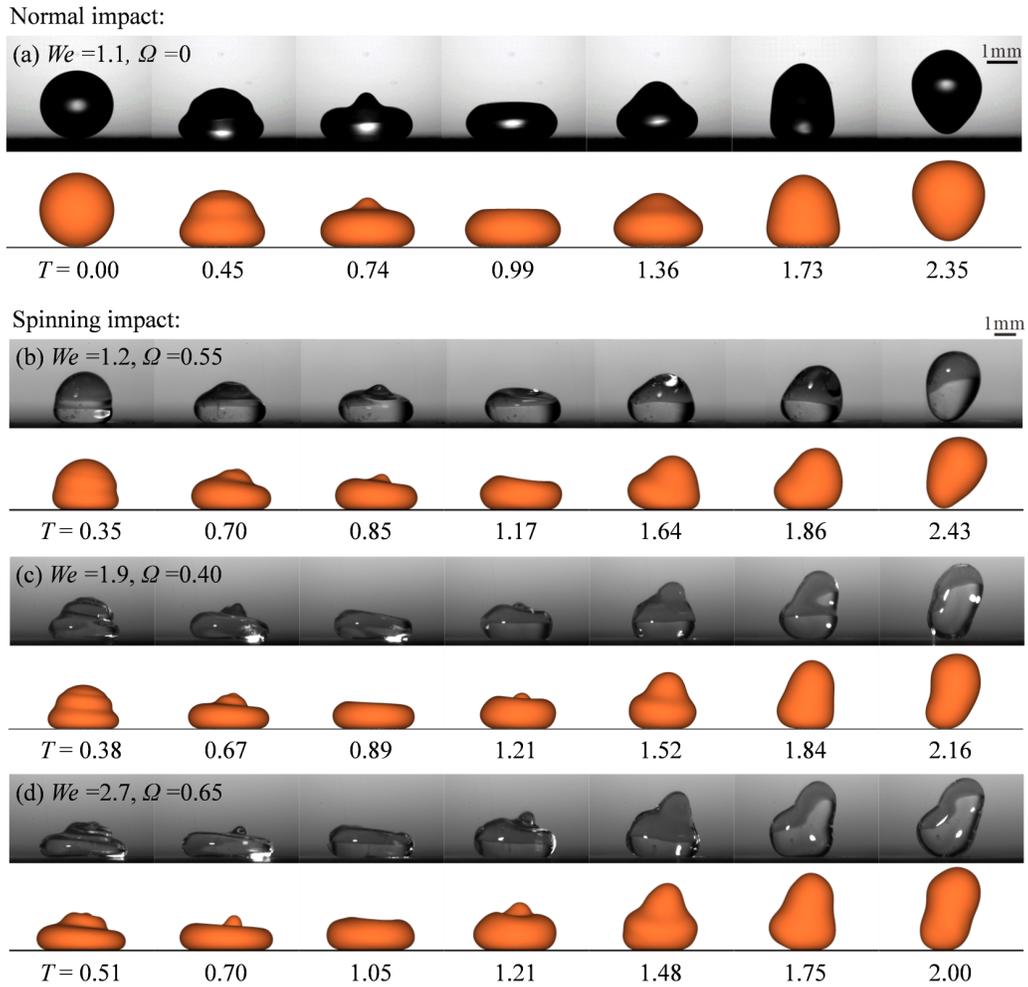

Figure 3: Comparison between the experimental and numerical results for droplets impacting onto a superhydrophobic surface: (a) $We$=1.1, $\Omega$ =0; (b) $We$=1.2, $\Omega$ =0.55; (c) $We$=1.9, $\Omega$ =0.40; (d) $We$=2.7, $\Omega$ =0.65.

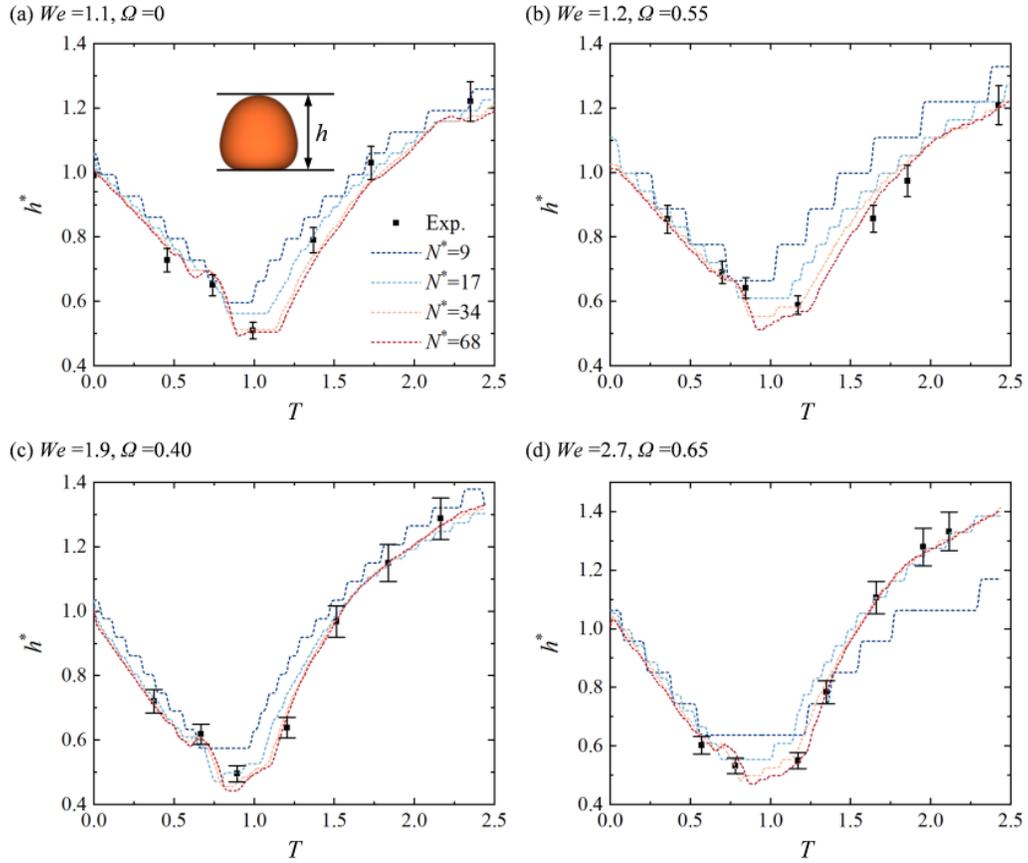

Figure 4: Grid independence test: (a) $We$=1.1, $\Omega$ =0; (b) $We$=1.2, $\Omega$ =0.55; (c) $We$=1.9, $\Omega$ =0.40; (d) $We$=2.7, $\Omega$ =0.65. $h^* = h/2R_0$ represents the normalized height of the droplet.

## 3.2. Tumbling rebound of spinning droplet

### 3.2.1. Typical scenarios of tumbling rebound

Fig. 5(a)∼(c) shows the temporal evolution of the normal impact process and two typical spinning impact processes. Compared to the symmetric spreading and rebound of normal impact in Fig. 5(a), two distinct types of droplet rebound were identified for the spinning impact. Specifically, in Fig. 5(b) with a lower normal and angular velocities of $We = 2$ and $\Omega = 1.13$, the droplet displays asymmetry during spreading and retraction, causing the surface wave and tip to deviate from the central axis at $T = 0.37$ and $T = 0.73$, spreading into a tilted disc shape at $T = 1.10$. Defining the positive x-axis as the front of the droplet and the negative x-axis as the rear, the droplet ultimately rebounds off the surface from the front side at approximately $T = 2.07$, which is shorter than the contact time of the normal rebound at around $T = 2.31$. As shown in Fig. 5(c), the asymmetric behavior becomes more pronounced at an increased angular velocity of $\Omega = 4.00$. The droplet spreads into a "long strip" shape at $T = 1.10$ and rebounds off the surface from the rear without retracting before $T = 1.83$. Consequently, we categorize the rebound process into two types of "front-raise tumbling rebound" and "rear-raise tumbling rebound", based on retraction lengths on both sides of the droplet. To accurately classify these two rebound types, we use the initial position of contact as the reference point, and employ $d_f(t)/D_0$ and $d_r(t)/D_0$ to represent the extended distances along the x-axis direction at the front and the rear of the droplet on the wall, respectively. Additionally, we define the retraction lengths from the maximum extended distance to the detachment point as the front side retraction length $\Delta d_f$ and the rear side retraction length $\Delta d_r$, both being normalized by the initial droplet diameter $D_0$. Fig.5(d) compares the temporal evolution of the extended distance for these two types of rebound. For the front-raise tumbling rebound, $\Delta d_f/D_0$ is greater than $\Delta d_r/D_0$, indicating that the droplet detaches from the front side of the surface first. The interpretation for the rear-raise tumbling rebound follows a similar logic, where the rear side lifts off earlier.

Fig. 6 shows the velocity and pressure fields in the central x-z plane for both the front-raise tumbling rebound (Fig. 6(a)) and the rear-raise tumbling rebound (Fig. 6(b)). The dimensionless pressure is defined as $p^* = pR_0/\sigma$, where $\sigma/R_0$ represents the capillary pressure. The flow direction is given by the vectors, and the magnitude is represented by the dimensionless velocity $V^* = \|V\|/\sqrt{\sigma/\rho_l R_0}$, where $\sqrt{\sigma/\rho_l R_0}$ is the capillary velocity. The vector superposition of vertical and angular velocities disturbs the symmetry of the droplet's internal flow field. As shown in Fig. 6(a), for the front-raise tumbling rebound, the vertical velocity on the front side of the droplet is higher, while that on the rear side is significantly lower at $T = 0.05$. The front side spreads more extensively under a larger impact inertial force, accompanied by a thinner thickness and a larger curvature, which results in a higher Laplace pressure at the edge of the droplet front side, as shown at $T = 0.54$. At $T = 0.97$, driven by the forward horizontal velocity component, a crater forms at the front side of the droplet, retracting with an upward and forward velocity induced by surface tension. Such asymmetric retraction of edge and crater causes an earlier detachment from the front side at $T = 1.85$. This also elucidates the reverse of droplet rotation reported in the experiments by Jia *et al.* [48] on the rebound of successive droplets obliquely impacting superhydrophobic surfaces. Such alteration is not a random event as they initially conjectured, but rather a consequence of the asymmetric capillary force. For the rear-raise tumbling rebound, as shown at $T = 0.05$ in Fig. 6(b), the increased angular velocity leads to a more pronounced asymmetry in the pressure and velocity distributions. Driven by both the initial spinning motion and the impact inertia, the droplet is stretched into a "long strip" shape without noticeable surface-tension-induced retraction before detachment, as observed at $T = 0.34$ and $T = 0.97$. Velocity field analysis further reveals that under significant rotational inertia, the rear side of the droplet lifts off earlier than the front side, which are governed by surface tension as previously described. The boundary between these two types of tumbling rebound will be analyzed in detail later.

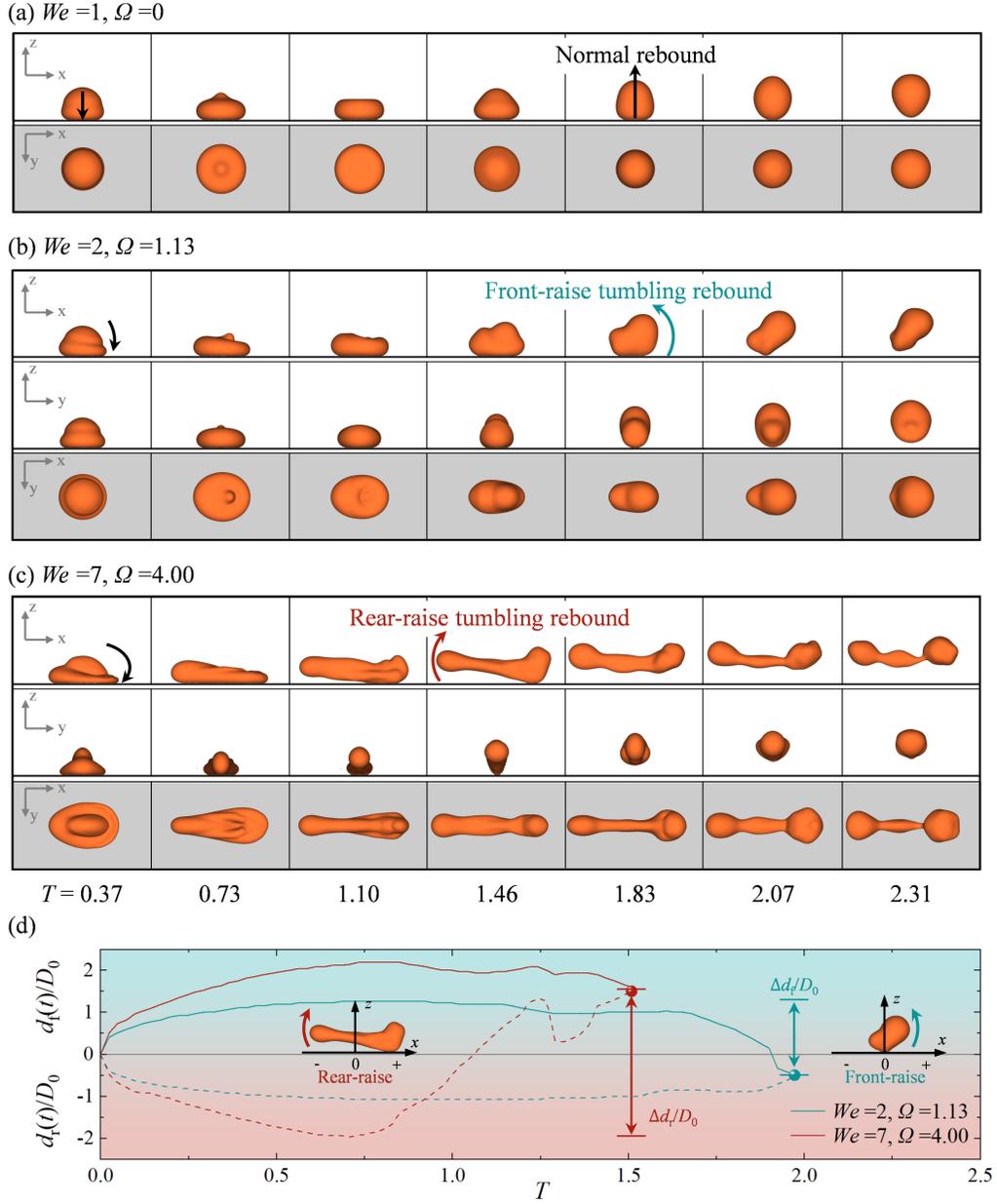

Figure 5: Temporal evolution of the droplet impact process: (a) $We=1$, $\Omega=0$ (normal rebound); (b) $We=2$, $\Omega=1.13$ (front-raise tumbling rebound); (c) $We=7$, $\Omega=4.00$ (rear-raise tumbling rebound). The top row shows the front view, the middle row shows the side view, and the bottom row shows the top view. Snapshots in the same column have the same dimensionless time $T$. (d) Temporal variation of the extended distances of the droplet's bottom surface along the x-axis direction. The solid and dashed lines indicate the positions of the droplet front edge (positive x-axis direction $d_f(t)/D_0$) and rear edge (negative x-axis direction $d_r(t)/D_0$), respectively. Solid dots represent the detachment moment of the droplet.

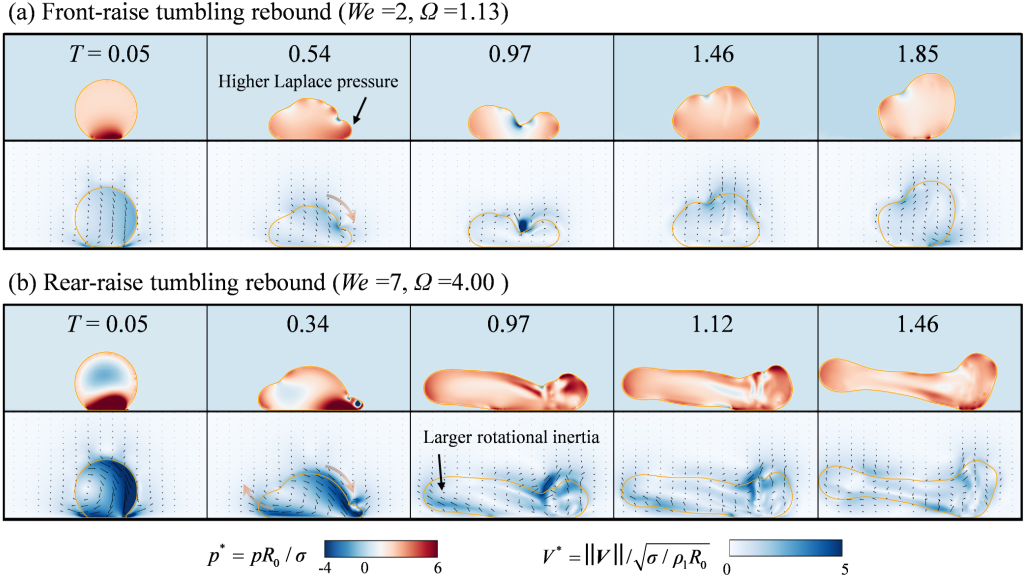

Figure 6: Velocity-pressure field distributions for two typical types of tumbling rebound: (a) front-raise tumbling rebound at $We=2$, $\Omega =1.13$; (b) rear-raise tumbling rebound at $We=7$, $\Omega =4.00$.

To comprehensively show the influence of spinning on the droplet rebound process, extensive simulations were performed in a wide range of controlling parameters. As shown in Fig. 7, with the increase of $\Omega$, the type of droplet rebound transitions from normal rebound, to front-raise tumbling rebound, and then to rear-raise tumbling rebound. Notably, such transitions are highly independent of $We$. As interpreted above, the type of rebound is determined by the competition between the motion of two sides, of which the front side is influenced by the capillary force, while the rear side is governed by the rotational inertia force. The timescales of these two processes can be characterized by the inertial-capillary time $t_\sigma \sim \sqrt{\rho_l R_0^3/\sigma}$ and the characteristic rotation time $t_\omega \sim \pi/(2\omega_0)$, respectively. For the front-raise tumbling rebound, a smaller $\Omega$ indicates that $t_\omega$ is greater than $t_\sigma$, and the droplet impact process remains similar to the classical vertical impact process. Consequently, the front side retracts and lifts first under the higher Laplace pressure. In contrast, for the rear-raise tumbling rebound, an increased $\Omega$ leads to a decrease in $t_\omega$, thereby transitioning into the rotational inertia regime, where the enhanced rotational inertial force lifts the rear side of the droplet first. Thus, a balance between $t_\omega$ and $t_\sigma$, with the boundary being independent

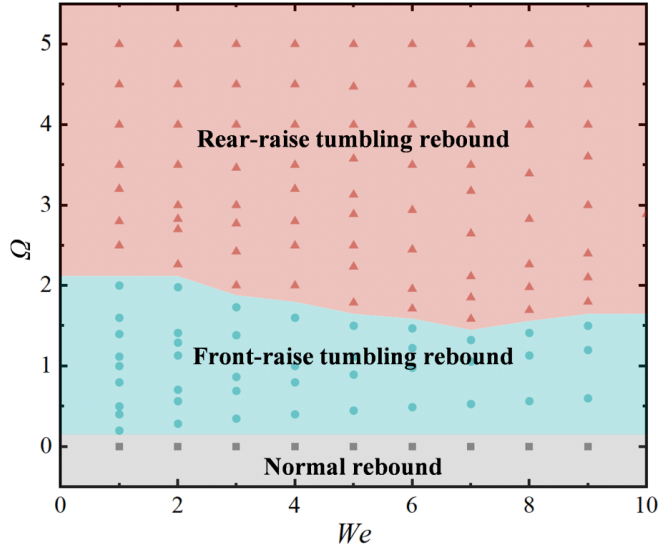

Figure 7: Regime map of the types of rebound for droplets impacting onto superhydrophobic surfaces at different $We$ and $\Omega$. The black squares, green circles, and red triangles correspond to normal rebound, front-raise tumbling rebound, and rear-raise tumbling rebound, respectively.

of $We$, can be defined as the transition line between front- and rear-raise tumbling rebounds.

*3.2.2. Evolution of angular momentum during tumbling rebound*

Since the two types of tumbling rebound show visually opposite spinning motions upon droplet detachment, one may naturally question whether the angular momentum of the droplet undergoes reversal. To address this issue, we look into the temporal variation of the angular momentum during the droplet impact process. The position vectors of the mass center $\mathbf{x}_{\mathrm{cm}}$ for the droplet can be calculated by:

$$\mathbf{x}_{\mathrm{cm}} = \int_{V_{\mathrm{d}}} \rho_{\mathrm{l}} \mathbf{x}_{\mathrm{l}} dV / m_{\mathrm{l}}, \tag{6}$$

where $V$ denotes volume, $V_d$ is the total volume of the droplet, $\mathbf{x}_l$ is the spatially varying position vector within the droplet, $m_l$ is the mass of the droplet. Then, the droplet angular momentum $\boldsymbol{L}(t)$ can be expressed as:

$$\boldsymbol{L}(t) = \int_{V_{\mathrm{d}}} \rho_{\mathrm{l}} (\mathbf{x}_{\mathrm{l}} - \mathbf{x}_{\mathrm{cm}}) \times \mathbf{u} dV_{\mathrm{d}}. \tag{7}$$

For the spinning droplet with an initial angular momentum $\boldsymbol{L}_0$ impacting on the surface and rebounds with a detachment angular momentum $\boldsymbol{L}_c$, the variation in the angular momentum can be expressed as:

$$\Delta \boldsymbol{L} = \boldsymbol{L}_c - \boldsymbol{L}_0 = \int_{t_c} \boldsymbol{M}(t)dt, \tag{8}$$

where $t_c$ is the surface-droplet contact time, $\boldsymbol{M}(t)$ is the torque generated by the reaction force $\boldsymbol{F}(t)$ from the air layer, which can be calculated as:

$$\boldsymbol{M}(t) = -\int_A (\mathbf{x}_1 - \mathbf{x}_{cm})d\boldsymbol{F}(t) \tag{9}$$

The negative sign indicates that the $\boldsymbol{M}(t)$ acts in the opposite direction of $\boldsymbol{L}_0$, and $A$ represents the area of the substrate surface. The normal reaction force $\boldsymbol{F}(t)$ on this surface can be further expressed as [78]:

$$d\boldsymbol{F}(t) = [(p - p_0)(\boldsymbol{I} \cdot \hat{\boldsymbol{z}}) - 2\mu_g(\boldsymbol{D} \cdot \hat{\boldsymbol{z}})] dA \tag{10}$$

where $p$ and $p_0$ denote the dynamic pressure distribution at the substrate surface and the ambient pressure, respectively, $\boldsymbol{I}$ represents the second-order identity tensor, $\hat{\boldsymbol{z}}$ is the upward-pointing unit vector normal to the substrate, $2\mu_g(\boldsymbol{D} \cdot \hat{\boldsymbol{z}})$ corresponds to the normal viscous force arising from the air layer between the droplet and the surface, which is negligible compared to the pressure. Therefore, the normal reaction force $\boldsymbol{F}(t)$ can be simply calculated as:

$$d\boldsymbol{F}(t) = dF(t)\hat{\boldsymbol{z}} = [(p - p_0)dA]\,\hat{\boldsymbol{z}} \tag{11}$$

Moreover, the potential torque arising from the viscous force within the air layer is neglected here, which will be justified later by comparing the magnitudes of viscous and inertial forces. Fig. 8 shows the angular momentum of the droplet at detachment from the surface under different $We$ and $\Omega$, which is normalized by the initial angular momentum $\boldsymbol{L}_0$. As $We$ and $\Omega$ increase, $\boldsymbol{L}_c/\boldsymbol{L}_0$ exhibits an evident trend of decline. A counterintuitive result is that the reversal of $\boldsymbol{L}_c/\boldsymbol{L}_0$ only occurs at higher $We$ and $\Omega$, contradicting the visually observed detachment spinning direction in Figs. 5(b) and (c).

To further explain this counterintuitive result, we analyze the temporal evolution of the torque $\boldsymbol{M}(t)$ exerted by the reaction force from the surface and the dimensionless angular momentum $\boldsymbol{L}(t)/\boldsymbol{L}_0$ of the droplet. Fig 9(a) shows that $\boldsymbol{M}(t)$ reaches its peak during the spreading phase, with its maximum value increasing as $\varOmega$ rises. This behavior is driven by the inertial following the droplet impact on the surface, as reflected in the temporal evolution of $\boldsymbol{L}(t)/\boldsymbol{L}_0$, where the declines observed for different $\varOmega$ cases during the spreading phase collapse onto a single curve. In the retraction phase, $\boldsymbol{M}(t)$ gradually decreases until the droplet fully detaches from the surface, at which point $\boldsymbol{M}(t)$ reaches zero and $\boldsymbol{L}(t)/\boldsymbol{L}_0$ stabilizes. Notably, during the front-raise tumbling rebound at a smaller $\varOmega$, a positive torque appears during the decrease in $\boldsymbol{M}(t)$, leading to a non-monotonic variation of "decrease-recovery" in $\boldsymbol{L}(t)/\boldsymbol{L}_0$. As $\varOmega$ increases, the rebound type transits to the rear-raise tumbling rebound, with $\boldsymbol{L}(t)/\boldsymbol{L}_0$ showing a continuous and significant decrease. Fig. 9(b) shows a similar temporal evolution of $\boldsymbol{M}(t)$ and $\boldsymbol{L}(t)/\boldsymbol{L}_0$ as in Fig. 9(a). As $We$ increases, both the peak and the overall value of $\boldsymbol{M}(t)$ increase, and $\boldsymbol{L}(t)/\boldsymbol{L}_0$ decreases more rapidly. The reason for these changes of $\boldsymbol{M}(t)$ at different $We$ and $\varOmega$ can be further explained through the spatial distributions of the reaction force $\boldsymbol{F}$ and torque $\boldsymbol{M}$ from the surface in Fig. 9. Fig. 10(a) presents a typical case of front-raise tumbling rebound at $We = 2, \varOmega = 0.28$. At lower $We$ and $\varOmega$, he impact inertia induces a relatively small asymmetric reaction force at $T = 0.19$ and $T = 0.34$, resulting in a slightly greater negative torque at the front side of the droplet (compared to the positive torque at the rear). As the front side of the droplet lifts, the reaction force of the front side decreases, leading to a gradual reduction in the negative torque. Consequently, net positive torques emerge at $T = 1.36$ and $T = 1.95$, explaining the temporal evolution of $\boldsymbol{M}(t)$ and the recovery of $L(t)/L_0$ observed in Fig. 10(a). Fig. 10(b) illustrates a typical case of front-raise tumbling rebound at a higher $\Omega = 4$, where the reaction force is significantly larger at the front side and smaller at the rear side of the droplet, resulting in a dominant negative torque over the positive torque. As the rear side lifts, the positive torque vanishes, leading $L(t)/L_0$ to continuously decrease without recovery. This explains the observed increase in $M(t)$ and the decrease in $L_c/L_0$ as $\Omega$ increases . Fig. 10(c) shows a case with an even higher $We = 9$, where the droplet exhibits a greater spreading distance, effectively extending the moment arm of the reaction force, thus amplifying $M$ by the enhanced asymmetry effect of the reaction force. Overall, the rear-raise tumbling rebound with consistent direction of visualized spinning motion with the initial droplet, experiences a more rapid decrease in $L(t)/L_0$ and hence being easier to results in $L_c/L_0$ reversal. The same reasoning applies to the non-reversal of $L_c/L_0$ in the front-raise tumbling rebound.

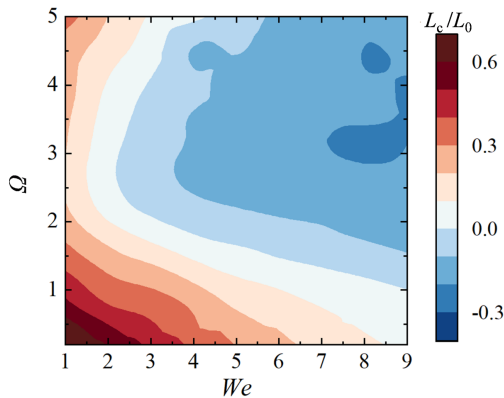

Figure 8: Regime map of the dimensionless angular momentum upon droplet detachment $\boldsymbol{L}_\mathrm{c}/\boldsymbol{L}_0$.

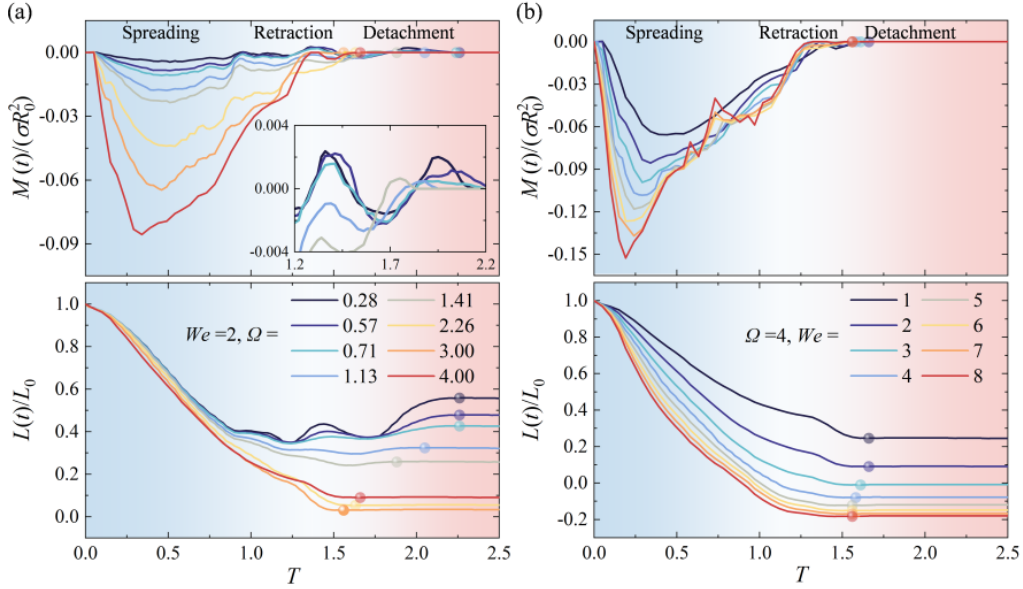

Figure 9: Temporal evolutions of the torque from surface $M(t)$ and the dimensionless angular momentum of the droplet $L(t)/L_0$: (a) at different $\Omega$ and fixed $We = 2$; (b) at different $We$ and fixed $\Omega = 4$. Solid dots represent the detachment moment of the droplet.

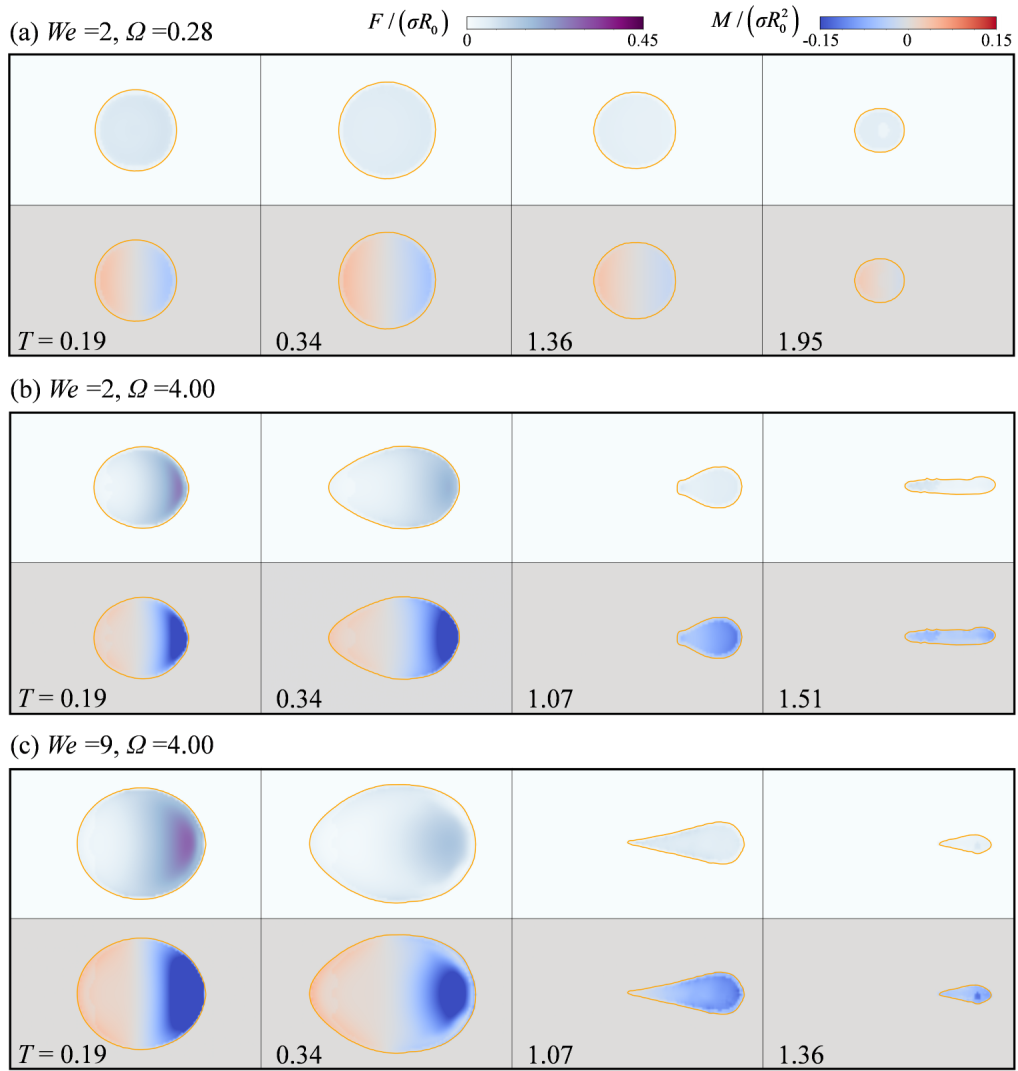

Figure 10: Field distributions of the reaction force $\boldsymbol{F}$ and the torque $\boldsymbol{M}$ during drop impact on the superhydrophobic surface: (a) $We = 2, \Omega = 0.28$; (b) $We = 2, \Omega = 4.00$; (c) $We = 9, \Omega = 4.00$.

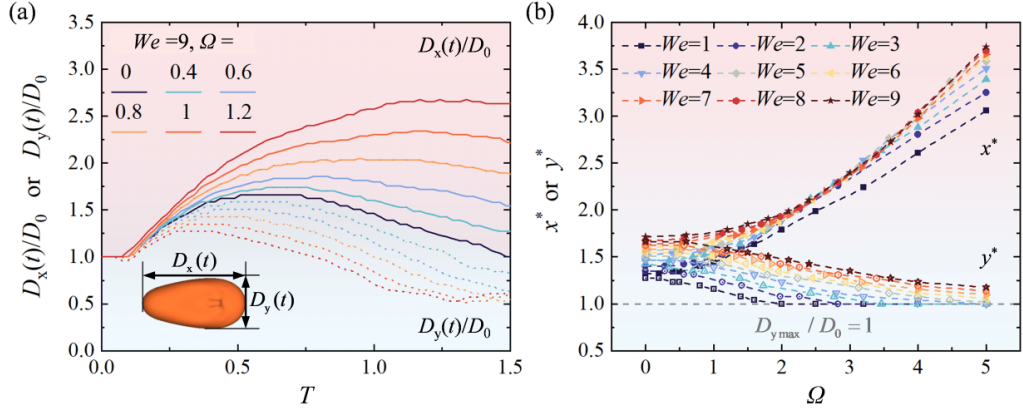

Figure 11: (a) Temporal variation of the droplet spreading lengths along the x-axis and y-axis under different $\Omega$ and a fixed $We = 9$. The black line represents the normal impact process, while the solid and dashed lines denote the long-axis $D_x(t)/D_0$ and short-axis $D_y(t)/D_0$ during the spinning impact process, respectively. (b) Variation of $x^*$ and $y^*$ with increasing $\Omega$ at different $We$. The solid and hollow markers denote $x^*$ and $y^*$, respectively, while the grey dashed line represents the geometric constraint of $y^*$.

### 3.3. Asymmetric deformation of spinning droplet

#### 3.3.1. Morphology of asymmetric deformation

Compared to the normal impact process, introducing spinning leads to significant asymmetric deformation for both types of tumbling rebound. Fig. 11(a) compares the temporal evolutions of the spreading distances along the x-direction $D_x(t)$ and along the y-direction $D_y(t)$ at a fixed $We = 9$, both of which are normalized by the initial droplet diameter $D_0$. For normal impact, as indicated by the black line, the variations in $D_x(t)/D_0$ and $D_y(t)/D_0$ are identical since the droplet spreads symmetrically. The overall trend shows a nonmonotonic variation of "increase-decrease", corresponding to the spreading and retraction phases, respectively. In contrast, during spinning impact, the droplet elongates and compresses along the x- and y-directions, and the difference between the long-axis $D_x(t)/D_0$ and the short-axis $D_y(t)/D_0$ enlarges with increasing $\Omega$, thus exhibiting a higher degree of asymmetry. Fig. 11(b) further compares the asymmetry of the droplet by using the maximum spreading lengths of $x^* = D_{xmax}/D_0$ and $y^* = D_{ymax}/D_0$. With the increase of $\Omega$, $x^*$ keeps increasing, while $y^*$ decreases until reaching the geometric constraint of $D_{ymax} \geq D_0$.

*3.3.2. Theoretical model*

The deformation and rebound of spinning droplet impact in both tumbling rebound types are affected by four forces: (i) the vertical inertial force $F_a = m_l V_0^2 / D_0 \sim O(10^{-4})$ N; (ii) the circumferential rotational inertia force $F_\omega = m_l \omega_0^2 R_0 \sim O(10^{-4})$ N; (iii) the capillary force $F_c = \sigma S / h \sim O(10^{-4})$ N, where $m_l$ is the mass of spinning droplet. $S$ and $h$ represent the spreading area and thickness of the droplet at the maximum spreading state, respectively; (iv) the viscous force $F_v$ between the droplet and the surface $(\mu_g \omega_0 R_0 / h_{air}) S \sim O(10^{-5})$ N, where the air layer thickness $h_{air}$ is on the order of micrometers according to the classical Landau Levich-Derjaguin law [30]. Therefore, among these four forces, the droplet impact process is primarily governed by the balance of $F_a$, $F_\omega$, and $F_c$.

As illustrated in Fig. 12(a) with the schematic of the spinning droplet impact process by asymmetric-velocity dumbbell model, the rotation effect can be abstracted as the rotational inertia force, which by coupling with the initial vertical inertial force, manifests as a vertical inertial force of varying magnitude along the x-axis. Consequently, an equivalent model applicable to both tumbling rebound types was established by replacing the original spinning impact with a modified normal impact, to separately describe the spreading dynamics on the front and rear sides of the droplet. On the front side, these two forces together generate an enhanced vertical downward inertial force $F_{x-f}$ to interact with the surface. This can be equivalently modeled as an intensified normal impact, yielding a maximum spreading radius on the front side of the droplet $R_{x-fmax}$ and a corresponding minimum thickness $h_{x-f\,min}$, through the balance between the capillary force $(\sigma / h_{x-fmin}^2)$ and the superimposed force of initial inertial force $(\rho_1 V_0^2 / D_0)$ and rotational inertia force $(\rho_1 \omega_0^2 R_0)$:

$$\frac{\sigma}{h_{x-f\,min}^2} \sim \rho_l \frac{V_0^2}{D_0} + \rho_l \omega_0^2 R_0, \tag{12}$$

Then, by coupling with the mass conservation relation in this equivalent model:

$$h_{x-fmin} R_{x-fmax}^2 \sim R_0^3, \tag{13}$$

$R_{x-fmax}$ could be derived by:

$$\frac{R_{x-f\,max}}{D_0} \sim \left(0.5 We + \Omega^2\right)^{1/4}. \tag{14}$$

As demonstrated by Fig. 12(b), $R_{x-f\,max}/D_0$ is in good agreement with Eq. (14) over a wide range of controlling parameters, which further validates the rationality of the equivalent model. On the rear side of the droplet, the initial inertial force opposes to the rotational inertia force, leading to a counteracting force $F_{\text{x-r}}$. When $\rho_l V_0^2/D_0 \geq \rho_l \omega_0^2 R_0$, the resultant force $F_{\text{x-r}} = \rho_l V_0^2/D_0 - \rho_l \omega_0^2 R_0$ acts as a downward inertial force. This allows for the establishment of a similar equivalent model as that on the front side, yielding a force balance between the counteracting inertial force and the capillary force. When $\rho_l V_0^2/D_0 < \rho_l \omega_0^2 R_0$, the resultant upward inertial force $F_{\text{x-r}} = \rho_l \omega_0^2 R_0 - \rho_l V_0^2/D_0$ stretches the droplet to facilitate spreading, thus maintaining an analogous force balance and spreading process:

$$\begin{cases} \dfrac{\sigma}{h_{\text{x-rmin}}^2} \sim \rho_l \dfrac{V_0^2}{D_0} - \rho_l \omega_0^2 R_0, & \rho_l \dfrac{V_0^2}{D_0} \geq \rho_l \omega_0^2 R_0 \\ \dfrac{\sigma}{h_{\text{x-rmin}}^2} \sim \rho_l \omega_0^2 R_0 - \rho_l \dfrac{V_0^2}{D_0}, & \rho_l \dfrac{V_0^2}{D_0} < \rho_l \omega_0^2 R_0 \end{cases}. \qquad (15)$$

Together with the conservation of mass for the maximum rear-side spreading radius $R_{\text{x-rmax}}$ and the corresponding minimum thickness $h_{\text{x-rmin}}$:

$$h_{\text{x-rmin}} R_{\text{x-rmax}}^2 \sim R_0^3. \qquad (16)$$

We have:

$$\begin{cases} \dfrac{R_{\text{x-rmax}}}{D_0} \sim (0.5We - \Omega^2)^{1/4}, & \rho_l \dfrac{V_0^2}{D_0} \geq \rho_l \omega_0^2 R_0 \\ \dfrac{R_{\text{x-rmax}}}{D_0} \sim (\Omega^2 - 0.5We)^{1/4}, & \rho_l \dfrac{V_0^2}{D_0} < \rho_l \omega_0^2 R_0 \end{cases}. \qquad (17)$$

Thus, the expression for $x^*$ can be further organized as:

$$\begin{aligned} x^* &= \dfrac{D_{\text{xmax}}}{D_0} = \dfrac{R_{\text{x-fmax}} + R_{\text{x-rmax}}}{D_0} \\ &\sim (0.5We + \Omega^2)^{1/4} + \mid 0.5We - \Omega^2 \mid^{1/4}, \end{aligned} \qquad (18)$$

which agrees well with the results over a wide range of control parameters, as shown in Fig. 12(c). For the short-axis spreading, the rotational effect elongates the long-axis of the droplet and hence compresses the short-axis under the constraint of mass conservation. The mass conservation during the spinning droplet impact process can be expressed as:

$$D_{\text{ymax}} D_{\text{xmax}} h_{\text{ymin}} \sim R_0^3, \qquad (19)$$

where $D_{\text{ymax}}$ is the maximum spreading diameter in the short-axis direction, and $h_y$ is the corresponding minimum thickness. The force balance along the short-axis between the capillary force and the initial inertial force is independent of the rotational inertia force:

$$\frac{\sigma}{h_{\text{ymin}}^2} \sim \rho_l \frac{V_0^2}{D_0}. \tag{20}$$

Therefore, $y^*$ can be derived from Eqs. (18), (19), and (20) as:

$$y^* = \frac{D_{\text{ymax}}}{D_0} \sim \frac{We^{1/2}}{(0.5We + \Omega^2)^{1/4} + \mid 0.5We - \Omega^2 \mid^{1/4}}, \tag{21}$$

which also agrees well with the numerical results, as shown in Fig. 12(d). We note that when $\Omega$ is large, results may deviate from Eq. (21) and collapse to $y^* = 1$, which is due to the geometric constraint $D_{\text{ymax}} \geq D_0$ described earlier. The expressions of $x^*$ and $y^*$ are controlled by both $We$ and $\Omega$, and broadly applicable to both the front-raise and rear-raise tumbling rebound of the droplet. Furthermore, when $\Omega = 0$, the expressions for $x^*$ and $y^*$ align with the maximum spreading ratio $D^* = D_m/D_0 \sim We^{1/4}$ of the classical normal impact process [14].

*3.4. Contact time of spinning droplet*

*3.4.1. Rapid detachment of spinning droplets*

In addition to asymmetric deformation, both types of tumbling rebound exhibit a rapid rebound phenomenon compared to normal impact. Fig. 13(a) illustrates the influence of $We$ and $\Omega$ on the dimensionless contact time $T_c = t_c/t_\sigma$. As $\Omega$ increases, $T_c$ significantly decreases to a certain threshold, while remaining relatively insensitive to the changes in $We$. More specifically, despite the differences observed at high $\Omega$ and low $\Omega$, $T_c$ shows a general trend of decrease with increasing $\Omega$, as shown in Fig. 13(b). When $\Omega$ is relatively low, the spinning impact process resembles a normal impact and results in $T_c$ being similar to those of typical normal impacts.

The rapid rebound phenomenon is closely related to the asymmetric deformation of the spreading droplet [17, 18, 79]. As illustrated at $T = 1.46$ in Fig. 4(c), the "long strip" shape of the droplet causes the rear side to lift off before the entire droplet detaches from the surface. In view of that the earlier detachment of the rear side reduces the effective contact with the surface during impact, we define the effective volume of the remaining droplet as $V_e \sim h_{\text{x-f min}} R_{\text{x-f max}} R_{\text{ymax}}$, which is smaller than the original volume of the droplet $V_d \sim R_0^3 \sim h_{\text{x-f min}} R_{\text{x-f max}}^2$. Then, considering the inertial-capillary time scale to be $t_\sigma \sim \sqrt{\rho_l R_0^3/\sigma} \sim \sqrt{\rho_l V_d/\sigma}$, we derive a dimensionless effective contact time as:

$$T_c = \frac{t_c}{t_\sigma} \sim \sqrt{\frac{V_e}{V_d}} \sim \sqrt{\frac{h_{\text{x-f min}} R_{\text{x-f max}} R_{\text{y max}}}{h_{\text{x-f min}} R_{\text{x-f max}}^2}} \sim \frac{R_{\text{y max}}}{R_{\text{x-f max}}}. \qquad (22)$$

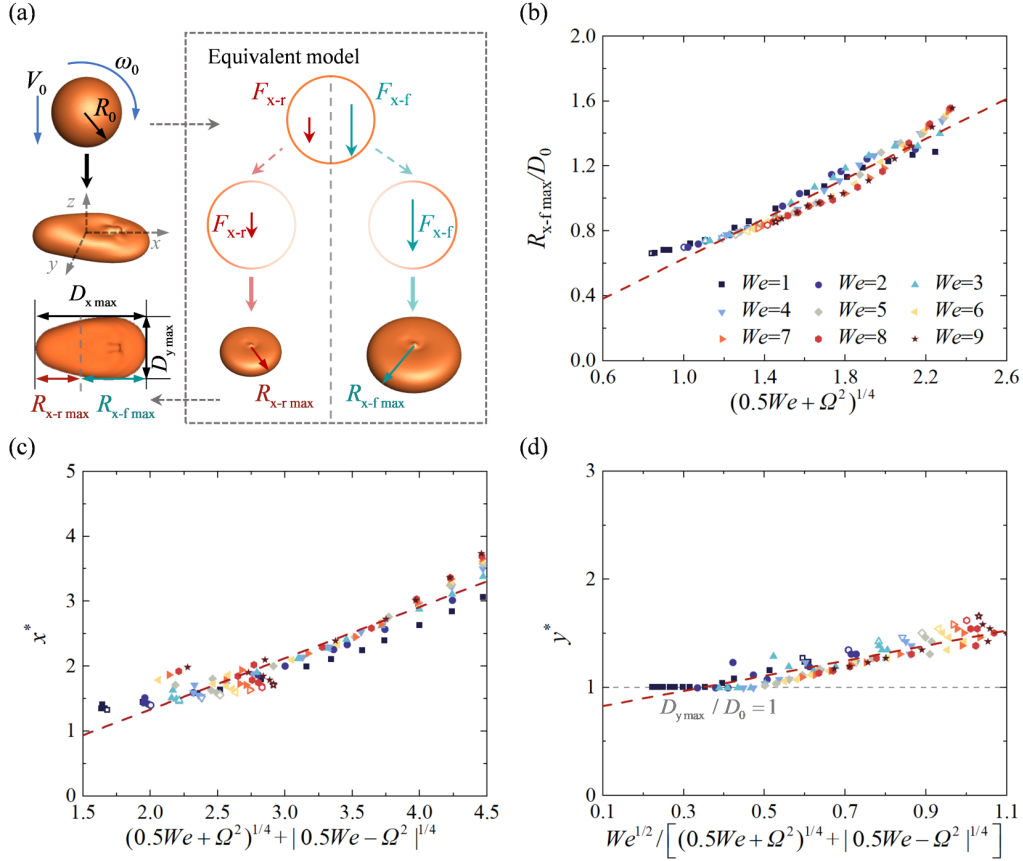

Figure 12: (a) Schematic of the spinning droplet impact process. (b) Relationship between $(0.5We + \Omega^2)^{1/4}$ and the dimensionless maximum spreading distance on the front side of the spinning droplet $R_{\text{x-fmax}}/D_0$. (c) Relationship between $(0.5We + \Omega^2)^{1/4}+ \mid 0.5We - \Omega^2 \mid^{1/4}$ and the dimensionless maximum spreading length along the long-axis $x^*$. (d) Relationship between $We^{1/2}/\left[(0.5We + \Omega^2)^{1/4}+ \mid 0.5We - \Omega^2 \mid^{1/4}\right]$ and the dimensionless maximum spreading length along the short-axis $y^*$. The gray dashed line represents the geometric constraint of $y^*$. The hollow markers represent the classical normal impact cases, and the solid markers represent the spinning impact cases. The red dotted lines in (b), (c), and (d) are the best fits based on Eqs (14), (18), and (21), with coefficients of $0.62 \pm 0.01, 0.79 \pm 0.02$, and $0.70 \pm 0.03$, respectively.

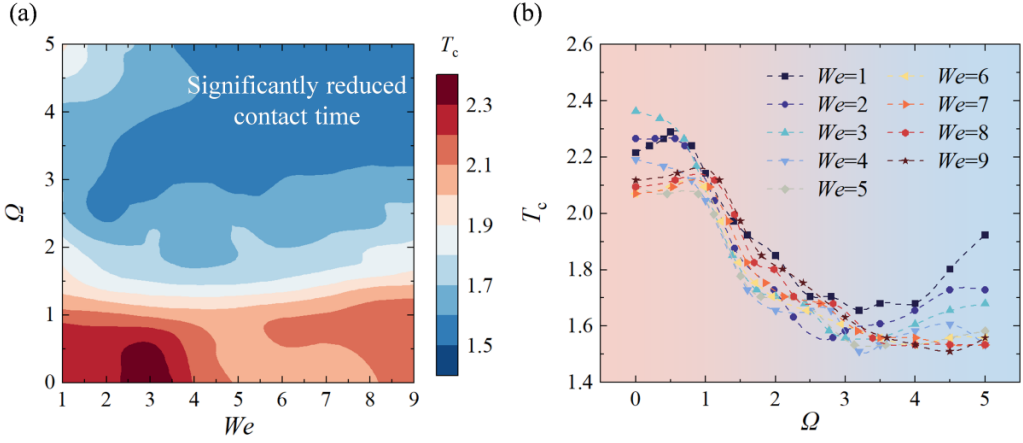

Figure 13: (a) Variation cloud of the dimensionless contact time $T_c = t_c/t_\sigma$ with different $We$ and $\Omega$. (b) Variation of $T_c$ as a function of $We$ and $\Omega$.

This relationship is evident in Fig. 14(a), where the data aligns well with Eq. (22). Combining Eqs. (14) and (21) further gives the expression for the dimensionless contact time as:

$$T_c \sim \frac{y^*}{R_{\text{x-f max}}/D_0} \\ \sim \frac{We^{1/2}}{\left[(0.5We+\Omega^2)^{1/4}+|0.5We-\Omega^2|^{1/4}\right](0.5We+\Omega^2)^{1/4}}, \qquad (23)$$

which is sufficiently validated from Fig. 14(b). When $\Omega = 0$, the theoretically derived expression for $T_c$ simplifies to a constant, corresponding to the contact time in the classical normal impact process [15]. Additionally, Eq. (23) could explain the deviation of results from the theoretical predictions at higher $\Omega$, as the geometric constraint limits $y^*$ to a minimum value of unity.

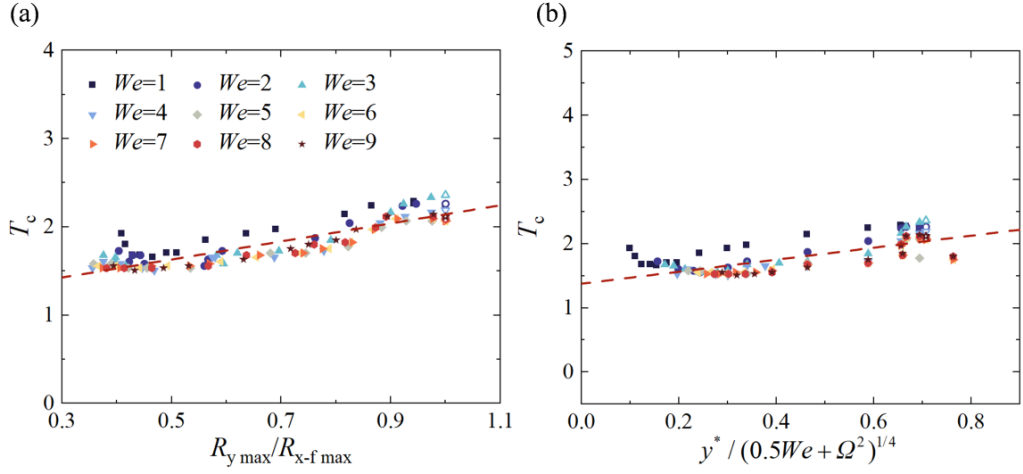

Figure 14: (a) Dimensionless contact time $T_c$ as a function of $R_{\text{ymax}}/R_{\text{x-fmax}}$. (b) $T_c$ as a function of $y^*/(0.5We+\Omega^2)^{1/4}$. The hollow markers represent the classical normal impact case. The dotted lines in (a) and (b) are the best fits based on Eqs. (22) and (23), with coefficients of $1.02 \pm 0.05$ and $1.38 \pm 0.04$, respectively.

3.4.2. Energy budget analysis

In addition to the asymmetric deformation, one may suspect whether the increased total energy of the droplet also contributes to the rapid rebound phenomena. Therefore, further analysis of the energy budget is conducted. Fig. 15 compares the temporal evolutions of different types of energy during the droplet impact process with the same total energy $E$ but different energy distribution ratios $e^* = E_\omega/E_{V_0}$, where $E_\omega$ and $E_{V_0}$ represent the kinetic energy from rotational and vertical impact velocities at the initial state, respectively. The energy budget includes the dimensionless kinetic energy $E_k/E$, the dimensionless surface energy $E_s/E$, the dimensionless gravitational potential energy $E_g/E$, and the dimensionless dissipative loss energy $E_d/E$, which can be represented as:

$$E_k = E_\omega + E_{V_0} = \int_{V_d} \frac{1}{2}\rho_l(\mathbf{u}\cdot\mathbf{u})dV_d, \tag{24}$$

$$E_s = \int_\Gamma \sigma d\Gamma, \tag{25}$$

$$E_g = \int_{V_d} \rho_l g z_l dV_d, \tag{26}$$

$$E_{\mathrm{d}} = \int_0^t \int_{V_{\mathrm{d}}} \Phi dV_{\mathrm{d}} dt. \tag{27}$$

where $g$ represents the acceleration of gravity, $\Gamma$ represents the area unit at the liquid-gas interface, and $\Phi$ is the viscous dissipation rate expressed as:

$$\begin{aligned}\Phi = \mu \bigg[ & 2\left(\frac{\partial u}{\partial x}\right)^2 + 2\left(\frac{\partial v}{\partial y}\right)^2 + 2\left(\frac{\partial w}{\partial z}\right)^2 \\ & + \left(\frac{\partial u}{\partial y}+\frac{\partial v}{\partial x}\right)^2 + \left(\frac{\partial v}{\partial z}+\frac{\partial w}{\partial y}\right)^2 + \left(\frac{\partial w}{\partial x}+\frac{\partial u}{\partial z}\right)^2 \bigg],\end{aligned} \tag{28}$$

where $u, v$ and $w$ represent the velocity components in the X, Y, and Z coordinate directions, respectively. The dashed line indicates the dimensionless contact time $T_{\mathrm{c}}$ for the complete detachment from the wall. As shown from Fig. 15(a) to Fig. 15(c), the contact time gradually decreases from $T_{\mathrm{c}} = 2.09$ and $T_{\mathrm{c}} = 1.58$ with the increases of $\Omega$ and $e^*$, suggesting that the contact time is reduced by the spinning motion rather than by the increased total energy of the droplet. As $e^*$ increases, the variations in $E_{\mathrm{g}}/E$ and $E_{\mathrm{d}}/E$ remain insignificant. The insignificant variation in $E_{\mathrm{d}}/E$ indicates that spinning does not yield an obvious increase in the energy loss. The increase in $e^*$ notably alters both $E_{\mathrm{k}}/E$ and $E_{\mathrm{s}}/E$, with their variations shifting from pronounced non-monotonic changes to a steady monotonic trend, corresponding to the variation of $\boldsymbol{L}(t)/\boldsymbol{L}_0$ during droplet spreading, retraction, and rebound. Such a transition facilitates the conversion between kinetic and surface energy, enabling the droplet to rebound rapidly from the surface.

## 4. Concluding remarks

The dynamics of spinning droplets impacting superhydrophobic surfaces were studied by using both high-speed imaging experiments and VOF simulations, focusing on tumbling rebound, asymmetric deformation, and contact time. The numerical methodology was quantitatively validated against experimental results, and extensive simulations were performed in wide ranges of normal impact $We$ number and dimensionless angular velocity $\Omega$. Results show that spinning motion induces two rebound types of front-raise tumbling rebound and rear-raise tumbling rebound. At low $\Omega$, asymmetric deformation generates higher Laplace pressure at the front side of droplet, causing it to retract first and leading to front-raise tumbling rebound. As $\Omega$ increases, enhanced rotational inertia lifts the rear side earlier, resulting in rear-raise tumbling rebound. A regime map of droplet rebound was established, with the critical $\Omega$ between front- and rear-raise tumbling rebound is a constant approximately independent of $We$, derived from comparing the inertial-capillary time and characteristic rotational time, indicating the transition from the vertical inertia regime to the rotational inertia regime. A counterintuitive finding is that the angular momentum of the spinning droplet upon detachment opposes the visually observed spinning direction of the tumbling rebound, due to the reversed torque induced by inertial impact. Moreover, a unified theoretical model was proposed to explain that increasing $\Omega$ enhances asymmetric deformation in both tumbling rebound types and consequently reduces the contact time of spinning droplets. The energy budget analysis further shows that the rapid rebound is a consequence of the spinning motion rather than the increased total energy of the droplet.

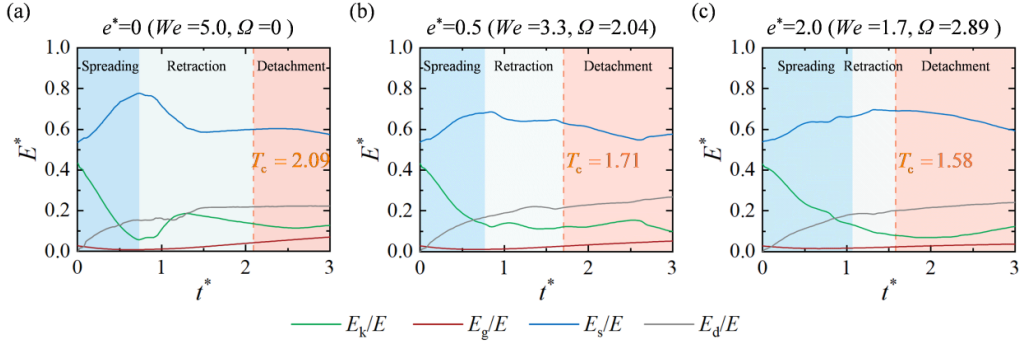

Figure 15: Energy budget of the droplet impact process: (a) $e^* = 0(We = 5, \Omega = 0)$; (b) $e^* = 0.5(We = 3.33, \Omega = 2.04)$; (c) $e^* = 2(We = 1.67, \Omega = 2.89)$. The dashed line indicates the dimensionless contact time. Droplets in these cases have an identical total energy $E$.

Leveraging emerging precision droplet manipulation techniques—such as acoustic levitation [80], optical tweezers and surface engineering [81] —future experimental investigations into more complex impact scenarios, including spinning droplet with high spin angular velocity, non-Newtonian effects and compound fluids, are well worth pursuing, as they provide broader applicability and practical relevance in both natural and industrial contexts.

**Acknowledgments**

This work was financially supported by the National Natural Science Foundation of China (No. 52076147). Peng Zhang acknowledges the support by the grants from the Research Grants Council of the Hong Kong Special Administrative Region, China (Project No. CityU 15222421 and CityU 15218820).

**Declaration of Interests**

The authors report no conflict of interest.

**CRediT author statement**

Jinyang Wang: Methodology, Investigation, Formal analysis, Writing - original draft. Feifei Jia: Investigation. Xiaoyun Peng: Investigation. Peng Zhang: Formal analysis, Funding acquisition, Writing - review & editing. Kai Sun: Conceptualization, Formal analysis, Funding acquisition, Writing - review & editing. Tianyou Wang: Supervision, Writing - review & editing.

**Data availability**

The data that support the findings of this study are available from the corresponding author upon reasonable request.

**References**

[1] M. R. O. Panao, A. L. N. Moreira, Intermittent spray cooling: A new technology for controlling surface temperature, International Journal of Heat and Fluid Flow 30 (1) (2009) 117–130.

[2] A. Moreira, A. Moita, M. Panao, Advances and challenges in explaining fuel spray impingement: How much of single droplet impact research is useful?, Progress in energy combustion science 36 (5) (2010) 554–580.

[3] H. Minemawari, T. Yamada, H. Matsui, J. Tsutsumi, S. Haas, R. Chiba, R. Kumai, T. Hasegawa, Inkjet printing of single-crystal films, Nature 475 (7356) (2011) 364–7.

[4] R. Rioboo, C. Tropea, M. Marengo, Outcomes from a drop impact on solid surfaces, Atomization and Sprays 11 (2) (2001) 155–165.

[5] A. L. Yarin, Drop impact dynamics: Splashing, spreading, receding, bouncing..., Annual Review of Fluid Mechanics 38 (1) (2006) 159–192.

[6] G. Liang, I. Mudawar, Review of drop impact on heated walls, International Journal of Heat and Mass Transfer 106 (2017) 103–126.

[7] M. Rein, Phenomena of liquid drop impact on solid and liquid surfaces, Fluid dynamics research 12 (2) (1993) 61–93.

[8] C. Josserand, S. T. Thoroddsen, Drop impact on a solid surface, Annual Review of Fluid Mechanics 48 (2016) 365–391.

[9] R. Blossey, Self-cleaning surfaces—virtual realities, Nature materials 2 (5) (2003) 301–306.

[10] M. J. Kreder, J. Alvarenga, P. Kim, J. Aizenberg, Design of anti-icing surfaces: smooth, textured or slippery?, Nature Reviews Materials 1 (1) (2016) 1–15.

[11] Z. Y. Ni, F. Q. Chu, S. K. Li, J. Luo, D. S. Wen, Impact-induced hole growth and liquid film dewetting on superhydrophobic surfaces, Physics of Fluids 33 (11) (2021) 112113.

[12] Z. Z. Zhang, B. Ge, X. H. Men, Y. Li, Mechanically durable, superhydrophobic coatings prepared by dual-layer method for anti-corrosion and self-cleaning, Colloids and Surfaces a-Physicochemical and Engineering Aspects 490 (2016) 182–188.

[13] N. Miljkovic, R. Enright, Y. Nam, K. Lopez, N. Dou, J. Sack, E. N. Wang, Jumping-droplet-enhanced condensation on scalable superhydrophobic nanostructured surfaces, Nano Lett 13 (1) (2013) 179–87.

[14] C. Clanet, C. BÉGuin, D. Richard, D. QuÉRÉ, Maximal deformation of an impacting drop, Journal of Fluid Mechanics 517 (2004) 199–208.

[15] D. Richard, C. Clanet, D. Quere, Contact time of a bouncing drop, Nature 417 (6891) (2002) 811.

[16] Y. Liu, L. Moevius, X. Xu, T. Qian, J. M. Yeomans, Z. Wang, Pancake bouncing on superhydrophobic surfaces, Nat Phys 10 (7) (2014) 515–519.

[17] J. C. Bird, R. Dhiman, H.-M. Kwon, K. K. Varanasi, Reducing the contact time of a bouncing drop, Nature 503 (7476) (2013) 385–388.

[18] Y. Liu, M. Andrew, J. Li, J. M. Yeomans, Z. Wang, Symmetry breaking in drop bouncing on curved surfaces, Nat Commun 6 (1) (2015) 10034.

[19] J. Du, Y. Li, X. Wu, Q. Min, Steerable directional bouncing and contact time reduction of impacting droplets on superhydrophobic stepped surfaces, J Colloid Interface Sci 629 (Pt A) (2023) 1032–1044.

[20] S. Yun, Bouncing of an ellipsoidal drop on a superhydrophobic surface, Sci Rep 7 (1) (2017) 17699.

[21] S. Yun, G. Lim, Ellipsoidal drop impact on a solid surface for rebound suppression, Journal of Fluid Mechanics 752 (2014) 266–281.

[22] N. Blanken, M. S. Saleem, C. Antonini, M. J. Thoraval, Rebound of self-lubricating compound drops, Science advances 6 (11) (2020) eaay3499.

[23] P. Gao, J. J. Feng, Spreading and breakup of a compound drop on a partially wetting substrate, Journal of Fluid Mechanics 682 (2011) 415–433.

[24] D. Liu, T. Tran, Emergence of two lamellas during impact of compound droplets, Applied Physics Letters 112 (20) (2018).

[25] S. Lin, B. Zhao, S. Zou, J. Guo, Z. Wei, L. Chen, Impact of viscous droplets on different wettable surfaces: Impact phenomena, the maximum spreading factor, spreading time and post-impact oscillation, J Colloid Interface Sci 516 (2018) 86–97.

[26] P. Sartori, E. Guglielmin, D. Ferraro, D. Filippi, A. Zaltron, M. Pierno, G. Mistura, Motion of newtonian drops deposited on liquid-impregnated surfaces induced by vertical vibrations, Journal of Fluid Mechanics 876 (2019).

[27] D. Bartolo, A. Boudaoud, G. Narcy, D. Bonn, Dynamics of non-newtonian droplets, Phys Rev Lett 99 (17) (2007) 174502.

[28] Y. H. Yeong, J. Burton, E. Loth, I. S. Bayer, Drop impact and rebound dynamics on an inclined superhydrophobic surface, Langmuir 30 (40) (2014) 12027–38.

[29] R. Zhang, P. Hao, F. He, Drop impact on oblique superhydrophobic surfaces with two-tier roughness, Langmuir 33 (14) (2017) 3556–3567.

[30] H. Zhan, C. Lu, C. Liu, Z. Wang, C. Lv, Y. Liu, Horizontal motion of a superhydrophobic substrate affects the drop bouncing dynamics, Phys Rev Lett 126 (23) (2021) 234503.

[31] H. B. Zhao, X. Han, J. Y. Li, W. Li, T. Huang, P. Yu, L. Q. Wang, Numerical investigation of a droplet impacting obliquely on a horizontal solid surface, Physical Review Fluids 7 (1) (2022) 013601.

[32] Š. Švikalo, C. Tropea, E. Ganić, Impact of droplets onto inclined surfaces, Journal of colloid and interface science 286 (2) (2005) 661–669.

[33] C. Antonini, F. Villa, M. Marengo, Oblique impacts of water drops onto hydrophobic and superhydrophobic surfaces: outcomes, timing, and rebound maps, Experiments in Fluids 55 (4) (2014).

[34] D. G. K. Aboud, A. M. Kietzig, Splashing threshold of oblique droplet impacts on surfaces of various wettability, Langmuir 31 (36) (2015) 10100–10111.

[35] C. Yin, T. Wang, Z. Che, M. Jia, K. Sun, Oblique impact of droplets on microstructured superhydrophobic surfaces, International Journal of Heat and Mass Transfer 123 (2018) 693–704.

[36] R. Tao, W. Fang, J. Wu, B. Dou, W. Xu, Z. Zheng, B. Li, Z. Wang, X. Feng, C. Hao, Rotating surfaces promote the shedding of droplets, Research 6 (2023) 0023.

[37] C. S. Park, H. Kim, H. C. Lim, Study of internal flow and evaporation characteristics inside a water droplet on a vertically vibrating hydrophobic surface, Experimental Thermal and Fluid Science 78 (2016) 112–123.

[38] K. Sun, L. Y. Shu, F. F. Jia, Z. Li, T. Y. Wang, Vibration-induced detachment of droplets on superhydrophobic surfaces, Physics of Fluids 34 (5) (2022) 11.

[39] P. Sartori, D. Quagliati, S. Varagnolo, M. Pierno, G. Mistura, F. Magaletti, C. M. Casciola, Drop motion induced by vertical vibrations, New Journal of Physics 17 (11) (2015).

[40] X. Noblin, R. Kofman, F. Celestini, Ratchetlike motion of a shaken drop, Phys Rev Lett 102 (19) (2009) 194504.

[41] W. Wang, C. Ji, F. Lin, J. Zou, S. Dorbolo, Water drops bouncing off vertically vibrating textured surfaces, Journal of Fluid Mechanics 876 (2019) 1041–1051.

[42] T. A. Duncombe, E. Y. Erdem, A. Shastry, R. Baskaran, K. F. Bohringer, Controlling liquid drops with texture ratchets, Adv Mater 24 (12) (2012) 1545–50.

[43] R. Clausius, Ueber die art der bewegung, welche wir wärme nennen, Annalen der Physik 176 (3) (2006) 353–380.

[44] Y. Jiang, A. Umemura, C. K. Law, An experimental investigation on the collision behaviour of hydrocarbon droplets, Journal of Fluid Mechanics 234 (1992) 171–190.

[45] K. L. Pan, K. K. L. Huang, W. T. Hsieh, C. R. Lu, Rotational separation after temporary coalescence in binary droplet collisions, Physical Review Fluids 4 (12) (2019) 123602.

[46] S. Bradley, C. Stow, Collisions between liquid drops, Philosophical Transactions of the Royal Society of London. Series A, Mathematical 287 (1349) (1978) 635–675.

[47] N. Ashgriz, J. Poo, Coalescence and separation in binary collisions of liquid drops, Journal of Fluid Mechanics 221 (1990) 183–204.

[48] Y. Jia, Z. Zhang, Y. Wang, S. Lin, Y. Jin, L. Chen, Successive rebounds of obliquely impinging water droplets on nanostructured superhydrophobic surfaces, Physics of Fluids 36 (7) (2024).

[49] C. He, Z. He, P. Zhang, Oblique bouncing of a droplet from a non-slip boundary: computational realization and application of self-spin droplets, International Journal of Multiphase Flow 167 (2023) 104548.

[50] S. Jung, M. K. Tiwari, N. V. Doan, D. Poulikakos, Mechanism of supercooled droplet freezing on surfaces, Nat Commun 3 (1) (2012) 615.

[51] B. Maneshian, K. Javadi, M. T. Rahni, R. Miller, Droplet dynamics in rotating flows, Advances in colloid and interface science 236 (2016) 63–82.

[52] R. J. Hill, L. Eaves, Nonaxisymmetric shapes of a magnetically levitated and spinning water droplet, Phys Rev Lett 101 (23) (2008) 234501.

[53] E. Janiaud, F. Elias, J. Bacri, V. Cabuil, R. Perzynski, Spinning ferrofluid microscopic droplets, Magnetohydrodynamics 36 (4) (2000) 301–314.

[54] F. Melo, J. F. Joanny, S. Fauve, Fingering instability of spinning drops, Phys Rev Lett 63 (18) (1989) 1958–1961.

[55] C. He, P. Zhang, Nonaxisymmetric flow characteristics in head-on collision of spinning droplets, Physical Review Fluids 5 (11) (2020) 113601.

[56] C. He, L. Yue, P. Zhang, Spin-affected reflexive and stretching separation of off-center droplet collision, Physical Review Fluids 7 (1) (2022) 013603.

[57] D. Richard, D. Quéré, Viscous drops rolling on a tilted non-wettable solid, Europhysics Letters (EPL) 48 (3) (1999) 286–291.

[58] A. F. W. Smith, K. Mahelona, S. C. Hendy, Rolling and slipping of droplets on superhydrophobic surfaces, Physical Review E 98 (3) (2018) 033113.

[59] Y. Lin, F. Chu, Q. Ma, X. Wu, Gyroscopic rotation of boiling droplets, Applied Physics Letters 118 (22) (2021).

[60] F. Jia, X. Peng, J. Wang, T. Wang, K. Sun, Marangoni-driven spreading of a droplet on a miscible thin liquid layer, J Colloid Interface Sci 658 (2024) 617–626.

[61] A. Bouillant, T. Mouterde, P. Bourrianne, A. Lagarde, C. Clanet, D. Quéré, Leidenfrost wheels, Nature Physics 14 (12) (2018) 1188–1192.

[62] M. Raffel, C. E. Willert, F. Scarano, C. J. Kähler, S. T. Wereley, J. Kompenhans, Particle image velocimetry: a practical guide, springer, 2018.

[63] J. Wei, J. Zhang, X. Cao, J. Huo, X. Huang, J. Zhang, Durable superhydrophobic coatings for prevention of rain attenuation of 5g/weather radomes, Nat Commun 14 (1) (2023) 2862.

[64] S. Popinet, collaborators, Basilisk c: volume of fluid method., http://basilisk.fr/ (2013-2025).

[65] P. Y. Lagrée, L. Staron, S. Popinet, The granular column collapse as a continuum: validity of a two-dimensional navier–stokes model with a µ(i)-rheology, Journal of Fluid Mechanics 686 (2011) 378–408.

[66] S. Popinet, Gerris: a tree-based adaptive solver for the incompressible euler equations in complex geometries, Journal of Computational Physics 190 (2) (2003) 572–600.

[67] S. Popinet, An accurate adaptive solver for surface-tension-driven interfacial flows, Journal of Computational Physics 228 (16) (2009) 5838–5866.

[68] M. Sussman, A. S. Almgren, J. B. Bell, P. Colella, L. H. Howell, M. L. Welcome, An adaptive level set approach for incompressible two-phase flows, Journal of Computational Physics 148 (1) (1999) 81–124.

[69] F. Gibou, L. Chen, D. Nguyen, S. Banerjee, A level set based sharp interface method for the multiphase incompressible navier–stokes equations with phase change, Journal of Computational Physics 222 (2) (2007) 536–555.

[70] R. Lebas, T. Menard, P. A. Beau, A. Berlemont, F. X. Demoulin, Numerical simulation of primary break-up and atomization: Dns and modelling study, International Journal of Multiphase Flow 35 (3) (2009) 247–260.

[71] J. U. Brackbill, D. B. Kothe, C. Zemach, A continuum method for modeling surface tension, Journal of computational physics 100 (2) (1992) 335–354.

[72] B. Zhang, V. Sanjay, S. Shi, Y. Zhao, C. Lv, X. Q. Feng, D. Lohse, Impact forces of water drops falling on superhydrophobic surfaces, Phys Rev Lett 129 (10) (2022) 104501.

[73] O. Ramírez-Soto, V. Sanjay, D. Lohse, J. T. Pham, D. Vollmer, Lifting a sessile oil drop from a superamphiphobic surface with an impacting one, Science advances 6 (34) (2020) eaba4330.

[74] J. De Ruiter, R. Lagraauw, D. Van Den Ende, F. Mugele, Wettability-independent bouncing on flat surfaces mediated by thin air films, Nature physics 11 (1) (2015) 48–53.

[75] G. Riboux, J. M. Gordillo, Experiments of drops impacting a smooth solid surface: a model of the critical impact speed for drop splashing, Physical review letters 113 (2) (2014) 024507.

[76] T. Young, Iii. an essay on the cohesion of fluids, Philosophical Transactions of the Royal Society of London 95 (1805) 65–87.

[77] J. A. van Hooft, S. Popinet, C. C. van Heerwaarden, S. J. A. van der Linden, S. R. de Roode, B. J. H. van de Wiel, Towards adaptive grids for atmospheric boundary-layer simulations, Boundary Layer Meteorol 167 (3) (2018) 421–443.

[78] L. D. Landau, E. M. Lifshitz, Fluid Mechanics: Volume 6, Vol. 6, Elsevier, 1987.

[79] A. Gauthier, S. Symon, C. Clanet, D. Quere, Water impacting on superhydrophobic macrotextures, Nat Commun 6 (1) (2015) 8001.

[80] S.-M. Argyri, L. Svenningsson, F. Guerroudj, D. Bernin, L. Evenäs, R. Bordes, Contact-free magnetic resonance imaging and spectroscopy with acoustic levitation, Nature Communications 16 (1) (2025) 3917.

[81] F. Wang, M. Liu, C. Liu, C. Huang, L. Zhang, A. Cui, Z. Hu, X. Du, Light control of droplets on photo-induced charged surfaces, National Science Review 10 (1) (2022).